**Magnetic behavior of Co ions in the exotic spin-chain compound, $Ca_3Co_2O_6$, from [59]Co NMR studies**

E.V. Sampathkumaran,[1,2,*] N. Fujiwara,[1] S. Rayaprol,[2] P.K. Madhu[2] and Y. Uwatoko[1]

[1]*The Institute for Solid State Physics, The University of Tokyo, 5-1-5 Kashiwanoha,*

*Kashiwa, Chiba 277 8581, Japan*

[2]*Tata Institute of Fundamental Research, Homi Bhabha Road, Mumbai 400005, India.*

## Abstract

We have performed field-swept [59]Co NMR measurements on the 'exotic' spin-chain material, $Ca_3Co_2O_6$, crystallizing in $K_4CdCl_6$-type rhombohedral structure, believed to exhibit two magnetic transition temperatures (around 24 and 12 K). This is the first NMR investigation of this family of compounds. We are able to detect the NMR signal below about 15 K, which is a conclusive proof for the existence of non-magnetic Co ions ($Co^{3+}$ - low-spin $3d^6$, in addition to magnetic Co ions) at least in a narrow temperature range around 10 K, thereby resolving a controversy on this issue in the literature. In addition, we find that the spin-lattice and spin-spin relaxation times undergo a dramatic increase below about 12 K, with a profound influence on the intensity of the NMR signal as a function of temperature, providing a microscopic evidence for the existence of 12K-transition. The spectral features present quite an interesting situation.

[*]**Corresponding author:** E-mail: sampath@tifr.res.in

(Revised on 28[th] April 2004)

# I. INTRODUCTION

The topics of low-dimensional magnetism and topological magnetic frustration have been attracting a lot of attention in recent years. In this respect, the family of compounds [1] of the type $(Sr,Ca)_3ABO_6$ containing magnetic ions A and B is notable, as the crystal structure ($K_4CdCl_6$-derived rhombohedra) provides an opportunity to study both the phenomena in a single family of compounds. In this article, we focus on $Ca_3Co_2O_6$, as this compound has aroused considerable level of experimental [2-12] and also of theoretical interest [13-17] among this family of compounds because of its peculiar magnetic characteristics as outlined below.

In this crystal structure, the chains made up of $AO_6$ (disorted) trigonal prisms and $BO_6$ (distorted) octahedra run parallel to the trigonal axis, with these polyhedra sharing a common face. The A/B ions form a triangular lattice in the basal plane [1] and the chains are separated by Ca/Sr ions. Generally, the inter-chain distance is about double of the intra-chain A-B distance. Till to date, none of these compounds has been found to be metallic. The compound of interest in this article is the only one in which both A and B sites can be fully occupied by the same ion. A calculation [17] suggests that this compound is a narrow-gap (0.086 eV) semi-conductor, though Co-Co distance within the chain is about 2.51 Å comparable to that in metallic cobalt. From the magnetism point of view, a transition around 24 K has been found arising from the antiferromagnetic ordering of the ferromagnetically coupled chains present at the vortices of the hexagon, while the third within the triangle in the basal plane remains incoherent. Therefore, this compound has been proposed [4] to serve  as an example for an uncommon magnetic structure, viz., 'partially disordered antiferromagnetic structure (PDA)', similar to the one seen in $CsCoCl_3$. As the temperature (T) is lowered, another transition has been reported in the close vicinity of 12 K.   However, the exact nature of this transition remains a matter of  dispute [4,6,7,10,12] with   some reports favoring the onset of  ferri-magnetic ordering involving the third chain, while some others offering evidence for  spin-glass freezing of the third chain.   Many other interesting properties have also been reported for this compound, for example, (i) observation [7,10,11] of a large frequency dependence of ac susceptibility, uncharacteristic of canonical spin-glasses, with an unusual magnetic-field (H) dependence [10],   (ii) several irreversible steps [4,7,10] in the isothermal magnetization (M) data at low temperatures, (iii) large thermopower [18-20] and (iv) T-dependent dimensional cross-over effects  in electrical conduction [9]. It is to be noted that, there are lively debates in the literature on the potential  application  of this compound or its derivatives [17-20] for thermoelectric devices.

For a compound of such a great current interest,  there is a fundamental controversy, namely,   on the spin/oxidation state of Co ions at the two sites. The bond strength concept [2] revealed that  Co-O length for B-site is smaller than that for A-site, on the basis of which it was proposed that the oxidation states of Co at both the sites are different (that is, larger and smaller than  3, respectively). The results of subsequent neutron diffraction studies [3] and spin-polarized  band structure calculations  [13] were not in agreement with this conclusion and are in favor of high-spin (HS) and low-spin (LS) $Co^{3+}$ states  ($d^6$ configuration) for A and B sites respectively. However, it is often stated (see, for instance, Refs.  12 and 15) that   indisputable experimental evidence is still absent for this scenario. The controversial nature of this issue is apparent from the fact that recent full-potential density-functional calculations [14] offer support to the proposal that Co at A and B sites are in HS $2^+$ and LS $4^+$ states respectively. Therefore, there is an urgent need to carry out more investigations to resolve this basic issue, as a deeper understanding (also theoretical [15]) of the underlying physics crucially depends on what the nature of the magnetic state of Co is at both the sites. With this primary motivation, we have undertaken field-swept $^{59}$Co (nuclear spin, I= 7/2) NMR investigations on this compound, the results of which are reported in this article. While we make other interesting observations, the results clearly reveal that non-magnetic Co ions are present, at least in a narrow temperature interval around 10 K, in this compound. We would like to mention that this is the first NMR investigation within this family of spin-chain compounds.

## II. EXPERIMENTAL DETAILS

The polycrystalline specimens employed in the present investigation were prepared by a solid-state reaction route [10] and characterized by x-ray diffraction to be single-phase materials. The samples were further characterized by magnetization measurements and the results obtained were found to be in good agreement with those reported in the literature [4]. $^{59}$Co NMR signal was searched by using a coherent pulsed spectrometer by sweeping the field at a frequency of 42 MHz at various temperatures down to 6 K; in addition, we performed the measurements at 24 MHz at 10 K.

## III. RESULTS AND DISCUSSION

We show in figure 1 the spectra obtained at 42 MHz. We would like to mention that the NMR signal could not be observed in the T-range 20 to 300 K (*vide-infra*). The central finding is that, as the T is lowered from 20 K to 15 K, some broad features start appearing. The spectra obtained for a given experimental conditions (specified in the figure caption) are shown in figure 1. As T is lowered to 10 K, there is a dramatic increase of the intensity with several well-resolved peaks. We are able to observe this signal, though with a reduction in intensity, at lower temperatures as well (Fig. 1) and we will return later in this article to a discussion of this intensity variation. It is quite well-known that the NMR signal from magnetically ordered Co ions should appear at frequencies as large as about 200 MHz in zero field (see, for instance, Refs. 21-23) due to large internal fields. The appearance of spectral features in the present NMR measurements (that is, at lower fields), therefore, establishes that non-magnetic Co ions are present in the compound.

The following questions remain to be answered: (1) What are these multiple peaks in Fig. 1 due to? (2) Is there any unpaired spin on the Co contributing to the above NMR signal? (3) Is there any indirect evidence from these spectra for the presence of magnetically ordered Co ions? In order to address these questions, we took NMR spectra (on a different specimen to verify the results) at a temperature at which the signal intensity is maximum (viz., 10 K) at one more frequency (24 MHz). The results are discussed below.

The most important observation (Fig. 2) is that the positions (in units of H) of all the intense peaks with respect to the reference position (given by gyromagnetic ratio 1.0054 MHz/kOe) are essentially unaltered with a change in NMR frequency (see Fig. 2). The observation that there is no *increase* of the *relative* spacings with *decreasing* frequency rules out any second-order quadrupolar effects [23], but consistent with a first-order quadrupolar effect; a quadrupolar interaction is expected as the symmetry of the sites of Co, having a non-zero electric quadrupolar moment, is non-cubic. However, one ideally expects only 7 peaks (3 satellites on either side of the central transition for I= 7/2), whereas there are more features appearing in the form of peaks or shoulders, for instance, for 42 MHz spectra below 40 kOe (see figure 1) in addition to a step around 50 kOe. This implies that the number of signals contributing to the spectra is more than one with different centers of gravity. We have observed a similar NMR pattern even if the sample was oriented in araldite at room temperature in the presence of a magnetic field of 70 kOe. We therefore tend to believe that the value of the internal hyperfine field is more than one, revealing a novel aspect of this material. It may be rewarding to focus future investigations to understand the implications of this finding. Possibly, ferri-magnetic and spin-glass regions coexist as inferred by viewing together all the results of neutron diffraction  [6] and bulk measurements [4,7,10,11], and the hyperfine field values are different for these regions.

We have also tried to simulate the spectra employing SIMPSON program [24] superimposing few sets of first-order quadrupolar patterns, by varying the parameters like line-width, quadrupolar coupling constants, asymmetry parameter and the resonance position. We find that it is very difficult to reproduce the experimental spectra, as it appears that there is a broad signal with unresolved quadrupolar satellites with a complicated line-shape, somewhat similar to that noted for another Co oxide in Ref. 21. The relative intensity of this broad signal is apparently less in 24 MHz-spectrum while compared to that in 42MHz-spectrum, as inferred from the depression of the shoulder around –4 kOe for the former in figure 2. There are field-induced changes in the isothermal M of the A-site Co ions, particularly a step-like feature around 35-40 kOe in M(H)    [4,7,9,10], which should in principle influence the spectra for 42 MHz, contributing to the difference in the shape of 24 MHz-spectrum in the range 0 to –5 kOe in figure 2.  In addition, there are H-induced changes in the electrical resistivity and the Coulomb gap [9] and it is not clear whether such effects have an effect on the relative intensities of the NMR peaks (and overall spectral-shape when H is swept over a wide range). Apparently, the spin-dynamics is very complex as revealed by ac susceptibility [7,10,11], as mentioned in the introduction.  Thus, it is not a trivial task to simulate the NMR line-shape of a quadrupolar nucleus in such a complicated quasi-one-dimensional magnetically ordered system.    Nevertheless, the main qualitative conclusion of this article is not hampered by this deficiency.  Though we do not base our conclusion on any such simulation, just for the benefit of the readers to understand the complexity of the NMR spectra and the interesting situation these present, we illustrate in figure 2 (bottom), how a simulated spectrum looks  if two quadrupolar sets are superimposed; the two-subspectra are denoted by *S1* and *S2* in the figure and the quadrupolar coupling constants and the asymmetry parameter are of  about 106 MHz and 1 respectively for both the sets; the central peaks of these sub-spectra are positioned at  -6.47 and –8.06 kOe respectively with respect to the reference point.

However, a large shift ($\Delta H$, more than 6 kOe) is transparent from the positions of the peaks even in the raw spectra. It is clear from the peak positions (figure 2) that the value of $\Delta H$ must be independent of the frequency. This finding clarifies the questions 2 and 3 raised above: (i) Intra-ionic paramagnetic contributions are negligible for the Co ions giving rise to these NMR spectra; if there are unpaired 3d electrons, one would expect that the paramagnetic shift caused by, say, core-polarization should have influenced [23] absolute positions of the peaks for different working frequencies, as $\Delta H/H$ (as defined in Ref. 23) should be a constant at a given temperature. This implies that Co responsible for this signal is in a trivalent state with a low-spin $3d^6$ configuration, which naturally [3] has to be the one at the octahedral site; (ii) The fact that $\Delta H$ is a fixed value implies that there is an internal field *transferred* from the *magnetically ordering* ions in the lattice (and not from any paramagnetic ion, as argued above), which in this case must be from the well-known magnetic ordering of the A-site Co ions. Thus, the observed spectra clearly render an evidence for the presence of both magnetic and non-magnetic ions.

Another point we make lies in the intensity variation of the NMR signal with a change in the temperature for the experimental conditions of the data shown in figure 1; in particular, a small change in T away from 10 K in figure 1 decreases the NMR signal intensity significantly. Since these Co ions are already in the low-spin state, it is rather difficult to propose a gradual change to a higher spin state (with a possible shift of the signal outside the field range of investigation) as a cause of the anomaly below 10 K. In order to understand the intensity variation, we measured spin-lattice relaxation time ($T_1$) and spin-spin relaxation time ($T_2$). It is obvious from figure 3 that both $T_1$ and $T_2$ (more sharply the former) undergo a dramatic increase below about 12 K, thereby microscopically establishing that there is a change in the magnetism of this compound around 12 K. It may be mentioned at this stage that the recovery curve of spin-echo looks as though there are two $T_1$ values, both falling in the range 2 to 7 msec, at this transition temperature, whereas unique $T_1$ value could be obtained at other temperatures. A careful comparison of the experimental conditions (mentioned in the caption of figure 1) and the values of these relaxation times provides an insight on the relative importance of these parameters in controlling NMR signal intensity in two temperature ranges (that is, above and below 10 K). For instance, *well below 10 K*, $T_2$ is larger than the pulse separation time, which rules out the role of $T_2$; also the increasing $T_2$ with decreasing T is not consistent with decreasing intensity. On the other hand, $T_1$ shows a remarkable increase from few msec at 10 K to about 400 msec at 7 K, thereby gradually getting much longer compared to the pulse repetition time with decreasing T. Therefore, it is $T_1$ variation that determines intensity below 10 K. However, *above 10 K*, the value of $T_1$ monotonically undergoes further fall (to 0.25 msec at 13.5 K), thus remaining shorter than the pulse repetition time; but the value of $T_2$ also gradually decreases with increasing T from about 460 μsec at 10 K to 14 μsec at 15 K, with the magnitudes comparable to the pulse separation above 10 K. Clearly, it is $T_2$ that is responsible for the intensity decrease above 10 K. This switchover of relative dominance of these relaxation times around 10 K must be closely associated with a corresponding change in the magnetism of this compound. Finally, it is not clear to us at present whether there is a change in the spin state of B-site Co (say, from the low spin to an intermediate spin configuration) above 10 K.

## IV. SUMMARY

To conclude, the present NMR results on $Ca_3Co_2O_6$ render evidence for the fact that the Co ions at both the trigonal-prismatic and octahedral sites are in the trivalent state (high- and low-spin electronic configurations respectively), at least in a narrow temperature interval around 10 K, thereby resolving this outstanding issue. Spin-latice and spin-spin relaxation times are found to exhibit a sudden increase below about 12 K, rendering a microscopic evidence for the existence of a transition below 12 K. Finally, we would like to emphasize that the spectral features are quite interesting, presumably carrying a lot of information related to the complexity of the  magnetism of this material, and therefore it would be rewarding  to focus future investigations on this aspect.

# References

1.  See, for instance, T.N. Nguyen, and H.C. zur Loye, J. Solid State Chem. **117,** 300 1995).

2.  H. Fjellvag, E. Gulbrandsen, A. Assland, A. Olsen and B.C. Hauback, J. Solid State Chem. 124, 190 (1996).

3.  S. Aasland, H. Fjellvag and B. Hauback, Solid State Commun. 101, 187 (1997).

4.  H. Kageyama, K. Yoshimura, K. Kosuge, H. Mitamura and T. Goto, J. Phys. Soc. Japan. 66, 1607 (1997).

5.  H. Kageyama, S. Kawasaki, K. Mibu, M. Takano, K. Yoshimura, and K. Kosuge, Phys. Rev. Lett. 79, 3258 (1997).

6.  H. Kageyama, K. Yoshimura, K. Kosuge, X. Xu and S. Kawano, J. Phys. Soc. Japan, 67, 357 (1998).

7.  A. Maignan, C. Michel, A.C. Masset, C. Martin and B. Raveau, Eur. Phys. J. B 15, 657 (2000).

8.  B. Martinez, V. Laukhin, M. Hernando, J. Fontcuberta, M. Parras and J.M. Gonzalez-Calbet, Phys. Rev. B 64, 012417 (2001).

9.  B. Raquet, M.N. Baibich, J.M. Broto, H. Rakoto, S. Lambert and A. Maignan, Phys. Rev. B 65, 104442 (2002).

10. S. Rayaprol, K. Sengupta and E.V. Sampathkumaran, Solid State Commun. 128, 79 (2003).

11. S. Rayaprol, K. Sengupta and E.V. Sampathkumaran, Proc. Indian. Acad. Sci. (Chem. Sci), 115, 553 (2003).

12.    V. Hardy, S. Lambert, M.R. Lees and D. McK Paul, Phys. Rev. B 68, 014424 (2003).

13.    M.-H. Whangbo, D. Dai, H.-J. Koo and S. Jobic, Solid State Commun. **125,** 413 (2003).

14.    R. Vidya, P. Ravindran, H. Fjellvag, A. Kjekshus and O. Eriksson, Phys. Rev. Lett. 91, 186404 (2003).

15.    R. Fresard, C. Laschinger, T. Kopp and V. Eyert, arXiv:cond-mat/0309031

16.    V. Eyert, C. laschinger, T. Kopp and R. Fresard, Chem. Phys. Lett. 385, 249 (2004).

17.    J. An and C.-W. Nan, Solid State Commun. **129**, 51 (2004).

18.    M. Mikami, R. Funahashi, M. Yoshimura, Y. Mori, and T. Sasaki, J. Appl. Phys. **94**, 6579 (2003).

19.    A. Maignan, S. Hebert, C. Martin, D. Flahaut, Mat. Sci. Eng. **104,** 121 (2003).

20.    K. Iwasaki, H. Yamane, S. Kubota, J. Takahashi and M. Shimada, J. Alloys and Compounds, **358**, 210  (2003).

21.    M. Itoh, Y. Nawaka, T. Kiyama, D. Akahoshi, N. Fujiwara and Y. Ueda, Physica B **329-333**, 751 (2003).

22.    H. Kubo, K. Zenmyo, M. Itoh, N. Nakayama, T. Mizota and Y. Ueda, J. Magn. Magn. Mater. (2004), in proof.

23.    See, for a review, *Metallic Shifts in NMR* by G.C. Garter, L.H. Bennett, and D.J. Kahan, in *Progress in Materials Science*, eds. B. Charmers, J,W. Christian and T.B. Massalski, **20** (Pergamon Press, Oxford, 1977).

24.    M. Bak, J.T. Rasmussen and N.C. Nielson, J. Magn. Res. **147,** 296 (2000).

Figure 1

[59]Co field-swept NMR spectra for the polycrystalline form of $Ca_3Co_2O_6$ at 15, 10, 9, 8, 7, and 6 K, obtained with a frequency of 42 MHz. The $\pi/2$-$\pi$ pulse separation time, the pulse repetition time and the pulse widths are 16μsec, 40 msec and 3 μsec respectively for these spectra. The dashed vertical line marks zero chemical shift position corresponding to the gyromagnetic ratio of 1.0054 MHz/kOe. The spectra at different temperatures are shifted in the vertical direction for the sake of clarity. The relative intensities of the spectra at various temperatures can be compared with each other.

Figure 2

(color online) The experimental [59]Co NMR spectra at 24 and 42 MHz taken at 10 K for $Ca_3Co_2O_6$ (on a different specimen, with a better signal to noise ratio) are shown. In the bottom figure, two first-order quadrupolar split sub-spectra (*S1*, *S2*) and their sum, simulated as described in the text are shown. *S1* and *S2* are shifted downwards along y-axis for the sake of clarity. The reference point (zero) on the x-axis is given by gyromagnetic ratio of 1.0054 MHz/kOe.

Figure 3

Temperature dependence of spin-lattice relaxation time ($T_1$) and spin-spin relaxation time ($T_2$) for $Ca_3Co_2O_6$, measured at the peak appearing at 32.4 kOe peak (in 42 MHz). The line through the data points serve as a guide to the eyes.

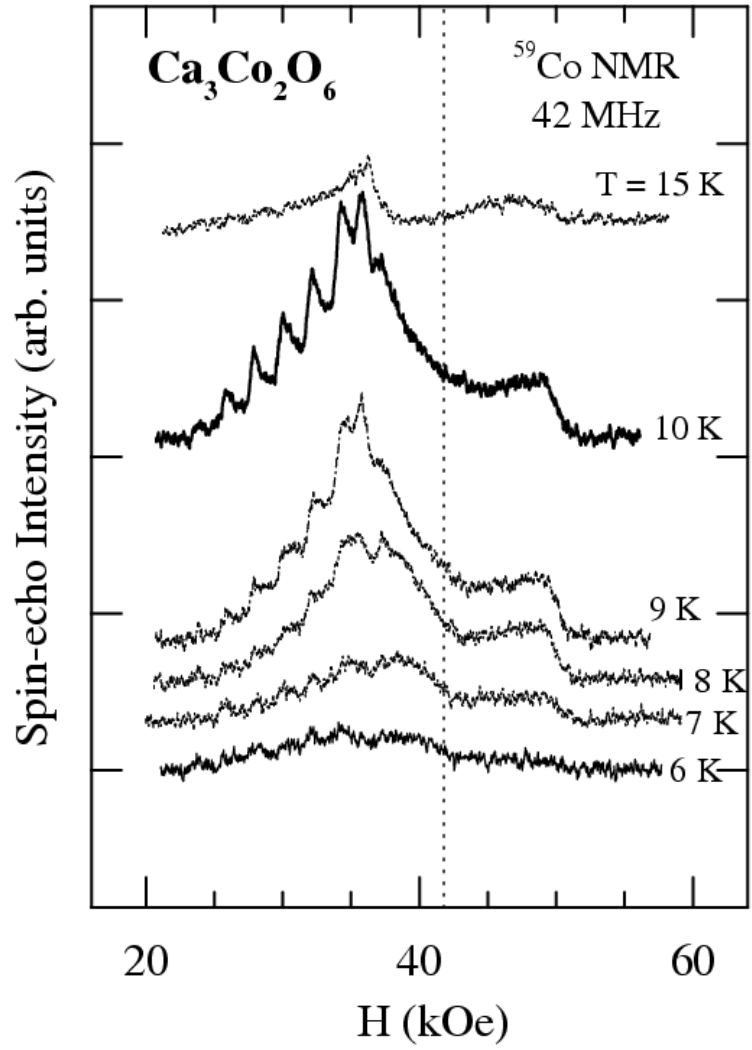

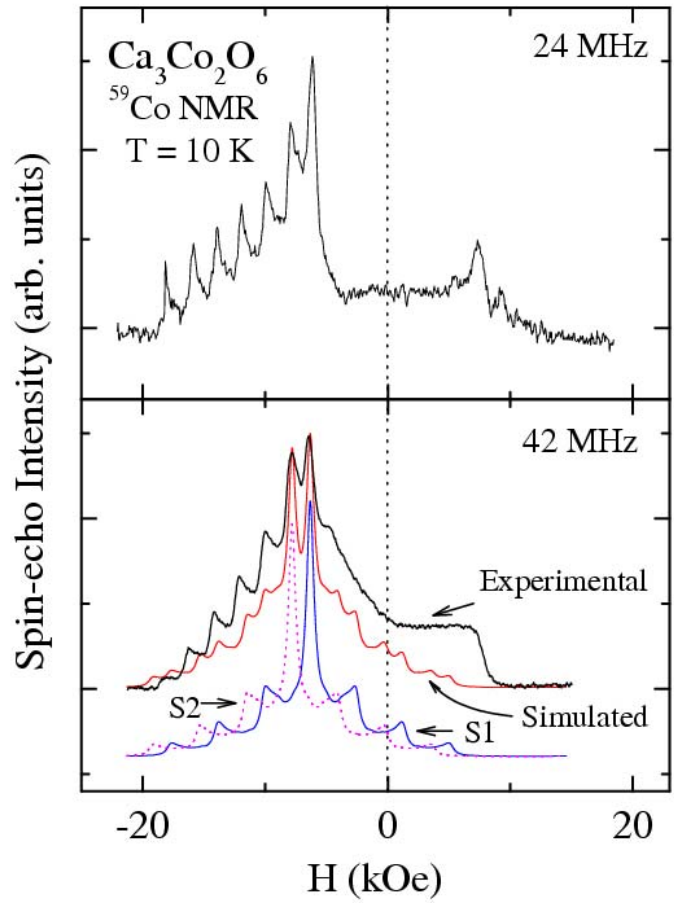

Figure 2

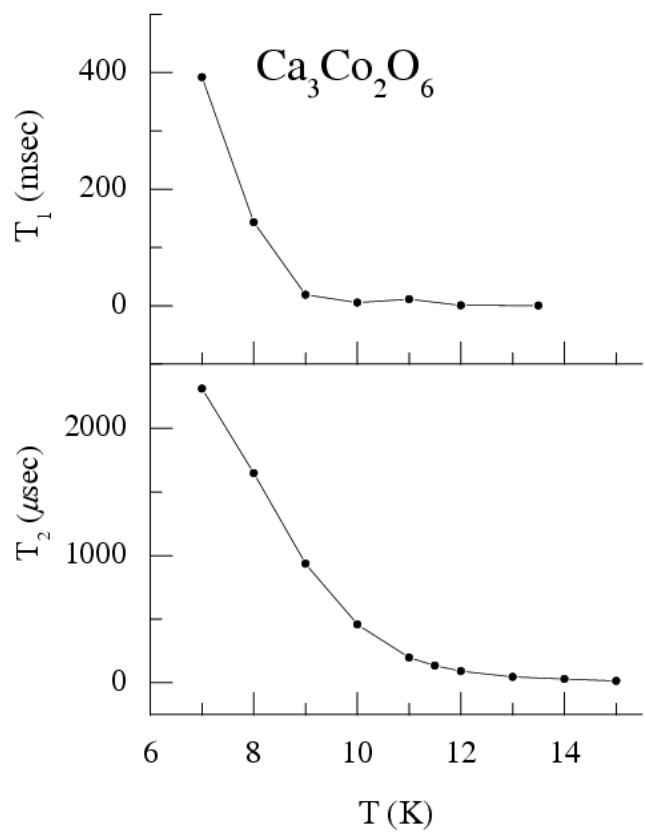

Figure 3